\documentclass[aps,preprint,amsmath,amssymb]{revtex4}
\usepackage{graphicx}

\begin{document}

\title
{\Large \bf
Equations of state for simple
liquids from the Gaussian equivalent representation method}

\author{
Dima Bolmatov\footnote{e-mail: bolmat@phys.nthu.edu.tw}}
\affiliation{Department of Physics, National Tsing Hua University,
Hsinchu 30013, Taiwan}


\begin{abstract}
 Within the framework of Gaussian equivalent representation method 
 a new procedure of obtaining equations of state for simple liquids is discussed in some technical details.
 The developed approach permits one to compute partition and distribution functions
 for simple liquids with arbitrary form of the central two-body
 potential of inter-molecular interaction. The proposed approach might become of great use  for computing thermodynamic and structural quantities of simple particle and polymer systems. We believe that this technique can also provide an interesting possibility to reduce the sign problem of other methods of computer simulation based on a functional integral approach. 
\end{abstract}

\maketitle

\section{Introduction}
 The calculation of partition and distribution functions is a basic problem of
statistical physics\cite{Hill-1}. All thermodynamics characteristics of
statistical systems are determined by these functions. As is well known, the calculation
of those quantities is a formidable problem\cite{Tson-1,Mat-1,Sc-1}.

In this paper we develop a functional integration method for 
systematic approximate calculations of classical partition functions of 
two-body potentials with positive and negative Fourier transforms over 
the entire density and temperature range.

 The Gaussian equivalent representation method has been recently introduced by
 Efimov and Ganbold in the context of quantum-field theory and statistical physics to
 compute integrals over Gaussian measure \cite{Ef-91,Ef-95,Ef-99}.
 The GER approach has already been proven to be very effective for computing thermodynamic properties and structural quantities of simple classical many-particle systems interacting with purely repulsive potentials like the Gauss-core or the Yukawa potential, possessing positive Fourier coefficients \cite{Ef-96,Nov-2006,Nov-2007} as well as for calculating the thermodynamics properties of flexible polymer systems \cite{Ba-1,Ba-2}. Moreover, it has successfully been employed to reduce the numerical sign problem in conjunction with Monte Carlo simulation \cite{Ba-3,Ba-4}. In the meantime,
 the real liquid description demands of considering the potentials
 having both attraction and repulsion parts \cite{HM-86,PABL-99}. In the present work the author extends the approach for systems, where the particles interact through potentials with positive and negative Fourier coefficients. This increases the range of applicability of this method for computational simulations \cite{Bae-1,Fre-1,Newman-1,Bae-2}.

 \section{The equations of state in the theory of simple liquids}

   The simple fluid of the particle density $n =N/V$ is thought of as a dense
   cloud of $N$ particles occupying the volume $V$ and interacting via two-body potential of the form
    $V({\bf x}-{\bf x}')$ \cite{FMN-81,BLSK-75}. Thermodynamics of such a system is described by the partition function
 \begin{equation}
  \label{e2.1}
  Z_V=\int_V\frac{d{\bf x}_1}{|V|}\dots\int_{V}\frac{d{\bf x}_N}{|V|}\exp[-\beta\sum_{i<j}^{N}
  V({\bf x}_i-{\bf x}_j)].
 \end{equation}
 For given two-body potential $V({\bf x}-{\bf x}')$ the free
 energy of the system can be computed and can be written in the form

  \begin{equation}
  \label{e2.1}
  E(n,\beta)= -\frac{1}{\beta} \lim_{|V| \rightarrow\infty} \ln{Z_{V}}.
  \end{equation}
 For systems of particles
 interacting via the two-body potentials having attraction $V_1$ and
 repulsion $V_2$ parts the total potential can conveniently be represented in the matrix
 form
 \begin{equation}
 \label{e2.3}
V^{-1}=\left( \begin{array}{cc}
            V_{1}^{-1} & 0 \\0 & V_{2}^{-1}  \end{array}   \right)
\end{equation}
 The differential operator $V^{-1}(\bold{x-y})$ satisfies the
 equation
\begin{equation}
\int d\bold{y}
V^{-1}(\bold{x-y})V(\bold{y-x^{\prime}})=\delta({\bf x}-{\bf x}'),\quad\mbox{\rm or}\quad \ V^{-1}V=I
\end{equation}
and  has a form
\begin{equation}
V^{-1}(\bold{x-y})= \frac{1}{\widetilde{V}(-\Delta)}
\delta(\bold{x-y})=\int\frac{d\bold{k}}{(2\pi)^{3}}\frac{e^{i\bold{k(x-y)}}}{\widetilde{V}(\bold{k})}
\end{equation}
where $\widetilde{V}(\bold{k})=\int d\bold{x} V(\bold{x})
e^{i\bold{kx}}$ is the Fourier-image of the potential.
The identity
\begin{equation}
\int\frac{D\phi}{C_{V}} e^{-\frac{1}{2}(\phi
V^{-1}\phi)+i(b\phi)}= e^{-\frac{1}{2}(bVb)}
\end{equation}
with
\begin{displaymath}
b(\bold{x})=\sqrt{\beta}\sum_{j=1}^{N}\delta(\bold{x-x_{j}})
\end{displaymath}
permits one to represent Boltzmann factor as
\begin{eqnarray}
\nonumber
&& e^{-\beta\sum_{i<j}^{N} V(\bold{x_{i}-x_{j}})}=
e^{N\frac{\beta}{2} V(0)-\frac{\beta}{2}\sum_{i,j}^{N}
V(\bold{x_{i}-x_{j}})}= \int\frac{D\phi}{C_{V}}
e^{-\frac{1}{2}(\phi V^{-1}\phi)+i(b\phi)+N\frac{\beta}{2} V(0)}=\\
&& \int\frac{D\phi}{C_{V}} e^{-\frac{1}{2} (\phi V^{-1}\phi)}
\prod_{j=1}^{N} e^{i
\sqrt{\beta}\phi(\bold{x_{j}})+\frac{\beta}{2}
V(0)}=\int\frac{D\phi}{C_{A}} e^{-\frac{1}{2}(\phi
V^{-1}\phi)}\prod_{j=1}^{N}\vdots
e^{i\sqrt{\beta}\phi(\bold{x_{j}})} \vdots_{V}
\end{eqnarray}
where the following notation is utilized
\begin{eqnarray}
(b\phi)&=&\int_{V} d\bold{x} b(\bold{x})\phi(\bold{x}),\\
(bVb)&=&\int_{V} \int_{V} d\bold{x} d\bold{y} b(\bold{x})
V(\bold{x-y}) b(\bold{y}),\\
(\phi V^{-1}\phi)&=&\int_{V}\int_{V} d\bold{x} d\bold{y}
\phi(\bold{x}) V^{-1}(\bold{x-y})\phi(\bold{y}).
\end{eqnarray}
Let us consider the partition function in the form of functional
integral
\begin{eqnarray}
\nonumber
Z_{V}&=&\int_{V}\frac{d\bold{x_{1}}}{|V|}\dots\int\frac{d\bold{x_{N}}}{|V|}
e^{-\beta\sum_{i<j}^{N} V_{1}(\bold{x_{i}-x_{j}})}
e^{\beta\sum_{i<j}^{N} V_{2}(\bold{x_{i}-x_{j}})}\\
\nonumber
&=&\frac{1}{|V|^{N}}\int\frac{D\phi}{\sqrt{detV_{1}}}\int\frac{D\psi}{\sqrt{detV_{2}}}
e^{-\frac{1}{2}(\phi V_{1}^{-1}\phi)} e^{-\frac{1}{2}(\psi
V_{2}^{-1}\psi)} \left[\int_{V} d\bold{x}\vdots
e^{i\sqrt{\beta}\phi(\bold{x})} e^{\sqrt{\beta}\psi(\bold{x})}
\vdots_{V} \right]^{N}\\
&=&\frac{1}{|V|^{N}}\int\frac{D\Phi}{\sqrt{detV}}
e^{-\frac{1}{2}(\Phi V^{-1}\Phi)} \left[ \int_{V} d\bold{x} \vdots
e^{i\sqrt{\beta}\Phi(\bold{x})J}\vdots_{V}
\right]^{N}=\frac{N!}{2\pi
i}\frac{1}{|V|^{N}}\oint\frac{dz}{z^{N+1}}I_{V}(z)
\end{eqnarray}
where
\begin{eqnarray}
\Phi(x)=(\phi(x),\psi(x)),\quad J=\left(\begin{array}{cc} 1\\-i
\end{array} \right),\quad
I_{V}(z)=\int\frac{D\Phi}{\sqrt{detV}} e^{-\frac{1}{2}(\Phi
V^{-1}\Phi)+z\int_{V} d\bold{x}\vdots e^{i\sqrt{\beta}\Phi
J}\vdots_{V}}.
\end{eqnarray}
 Following the line of argument of the GER, the integral $I_{V}(z)$ is written
 as follows
\begin{eqnarray}
I_{V}(z)&=&\sqrt{\frac{detD}{detV}}\int\frac{D\Phi}{\sqrt{detV}}
e^{-\frac{1}{2}(\Phi
D^{-1}\Phi)-\frac{1}{2}(\Phi[V^{-1}-D^{-1}]\Phi)-(\Phi
V^{-1}\Phi_{0})}\\
&\times& e^{z\int_{V}d\bold{x}
i\sqrt{\beta}(\Phi(\bold{x})+\Phi_{0})J+\frac{1}{2}(V_{1}(0)-V_{2}(0))}=
e^{W_{0}(z)}\int\frac{D\Phi}{\sqrt{detD}} e^{-\frac{1}{2}(\Phi
D^{-1}\Phi)} e^{W_{I}[\Phi]},
\end{eqnarray}
where
\begin{equation}
(JVJ)=\int d\bold{x}\int d\bold{y} J\delta(\bold{x-x^{\prime}})
V(\bold{x-y}) J\delta(\bold{x^{\prime}-y})= V_{1}(0)-V_{2}(0).
\end{equation}
As a result, we obtain two equations
\begin{eqnarray}
\nonumber
 \mbox{Equation I}: &\frac{1}{2}&\int d\bold{x}\int
d\bold{y}(\Phi(\bold{x})[V^{-1}(\bold{x-y})-D^{-1}(\bold{x-y})]|\Phi(\bold{y}))\\
+ &\frac{1}{2}&\int d\bold{x}\int d\bold{y}[-\beta(\Phi J)^{2}]
e^{i\sqrt{\beta}(\Phi_{0}J)+\frac{\beta}{2}[J(V-D)J]}=0
\end{eqnarray}
or, more specifically this equation reads
\begin{eqnarray}
\nonumber
\int\frac{d\bold{k}}{(2\pi)^{3}}\frac{e^{i\bold{k(x-y)}}}{\widetilde{V}(\bold{k})}&-&
\int\frac{d\bold{k}}{(2\pi)^{3}}\frac{e^{i\bold{k(x-y)}}}{\widetilde{D}(\bold{k})}
=-z\beta(J)^{2}e^{i\sqrt{\beta}(\Phi_{0}J)+\frac{\beta}{2}[J(V-D)J]}\Longrightarrow\\
\label{EQ-I}
\frac{1}{\widetilde{V}(\bold{k})}&-&\frac{1}{\widetilde{D}(\bold{k})}=-z\beta(J)^{2}e^{i\sqrt{\beta}(\Phi_{0}J)+
\frac{\beta}{2}[J(V-D)J]}.
\end{eqnarray}

\begin{eqnarray}
&& \mbox{Equation II}:\ -(\Phi V^{-1}\Phi_{0})+iz\sqrt{\beta}\int
d\bold{x}(\Phi(\bold{x})J)e^{i\sqrt{\beta}(\Phi_{0}J)+\frac{\beta}{2}[J(V-D)J]}
\end{eqnarray}
or
\begin{equation}
  \Phi_{0i}=iz\sqrt{\beta}\int
d\bold{x}(\Phi
J)e^{i\sqrt{\beta}(\Phi_{0}J)+\frac{\beta}{2}[J(V-D)J]},\ i=1,2.
\end{equation}
Assume $\Phi_{0i}=ic_{i}/\beta$, where
\begin{equation}
\label{EQ-II}
c_{i}=z\beta\widetilde{V}_{ij}(0)J_{j}e^{-c_{i}J_{i}+\frac{\beta}{2}[
J(V-D)J]}.
\end{equation}
Making use of (\ref{EQ-II}) in (\ref{EQ-I}), we obtain
\begin{eqnarray}
&& \left( \frac{1}{\tilde{V}(k)}- \frac{1}{\tilde{D}(k)}\right
)_{11}= - \frac{c_{1}}{\tilde{V_{1}}(0)} , \ \left(
\frac{1}{\tilde{V}(k)}- \frac{1}{\tilde{D}(k)}\right )_{22}=
\frac{c_{1}}{\tilde{V_{1}}(0)},\\
&& \left( \frac{1}{\tilde{V}(k)}- \frac{1}{\tilde{D}(k)}\right
)_{12}=\left( \frac{1}{\tilde{V}(k)}- \frac{1}{\tilde{D}(k)}\right
)_{21}= - \frac{c_{2}}{\tilde{V_{2}}(0)}
\end{eqnarray}
This can be combined in the matrix
\begin{eqnarray}
\frac{1}{\tilde{D}(k)}=\left( \begin{array}{cc}
\frac{1}{\tilde{V_{1}}(k)}+\frac{c_{1}}{\tilde{V_{1}}(0)} &
\frac{c_{2}}{\tilde{V_{2}}(0)} \\ \frac{c_{2}}{\tilde{V_{2}}(0)} &
\frac{1}{\tilde{V_{2}}(k)}-\frac{c_{1}}{\tilde{V_{1}}(0)}
\end{array} \right)=\tilde{D}^{-1}(k).
\end{eqnarray}
Thus, we  obtain for $\tilde{D}(k)$
\begin{equation}
\tilde{D}(k)=\frac{1}{\Delta}\left(
\begin{array}{cc}
\frac{1}{\tilde{V_{2}}(k)}-\frac{c_{1}}{\tilde{V_{1}}(0)} &
-\frac{c_{2}}{\tilde{V_{2}}(0)} \\ -\frac{c_{2}}{\tilde{V_{2}}(0)}
&
\frac{1}{\tilde{V_{1}}(k)}+\frac{c_{1}}{\tilde{V_{1}}(0)}\end{array}
\right),
\end{equation}
where
\begin{displaymath}
\Delta=det\left(
\frac{1}{\tilde{D}(k)}\right)=\left(\frac{1}{\tilde{V_{1}}(k)}+\frac{c_{1}}{\tilde{V_{1}}(k)}
\right)\left(\frac{1}{\tilde{V_{2}}(k)}-\frac{c_{1}}{\tilde{V_{1}}(0)}\right)
-\left(\frac{c_{2}}{\tilde{V_{2}}(0)}\right)^{2}.
\end{displaymath}
At large $N$ we can use Stirling's formula
\begin{equation}
\frac{N!}{|V|^{N}}\Longrightarrow
e^{|V|(n\ln{n}-n)}=e^{|V|f_{n}},\quad \mbox{\rm where}\quad f_{n}=n\ln{n}-n.
\end{equation}
It is convenient to introduce function
\begin{equation}
\label{func}
|V|R(c)=|V|f_{n}+W_{0}-N\ln{z}
\end{equation}
where
\begin{eqnarray}
\nonumber
W_{0}&=&\frac{1}{2}\ln{\frac{detD}{detV}}-\frac{1}{2}Tr[D(V^{-1}-D^{-1})]-\frac{1}{2}(\Phi_{0}V^{-1}\Phi_{0})
+z\int_{V}d\bold{x}
e^{-cJ+\frac{\beta}{2}(J[V-D]J)}\\ \nonumber
&=&\frac{1}{2}|V|\int\frac{d\bold{k}}{(2\pi)^{3}}\left[
\ln{\frac{\Delta^{-1}(\bold{k})}{\tilde{V_{1}}(\bold{k})\tilde{V_{2}}(\bold{k})}}+Tr(1-\tilde{D}(\bold{k})\tilde{V}^{-1}(\bold{k}))
\right]\\
 &-&\frac{\Phi_{01}^{2}}{2\tilde{V_{1}}(0)}-\frac{\Phi_{02}^{2}}{2\tilde{V_{2}}(0)}+\frac{c_{1}}{\beta\tilde{V_{1}}(0)}
\end{eqnarray}
and
\begin{equation}
\ln{z}=\ln{\frac{c_{1}}{\beta\tilde{V_{1}}(0)}}+cJ-\frac{\beta}{2}(J[V-D]J).
\end{equation}
After some algebra we obtain the partition function in the form:
\begin{equation}
Z_{V}=\frac{1}{2\pi
i}\oint\frac{dz}{z}e^{|V|R(c(z))}\int\frac{D\phi}{\sqrt{detD}}e^{-\frac{1}{2}(\Phi
D^{-1}\Phi)}e^{W_{I}[\Phi]},
\end{equation}
where
\begin{equation}
W_{I}[\Phi]=\frac{c(z)}{b}\int_{V}d\bold{x}\vdots
e_{2}^{i\sqrt{\beta}\Phi(\bold{x})}\vdots_{D}.
\end{equation}
The partition function integral has been derived in a similar from
in section III of reference \cite{Ba-5} for potential models with positive Fourier
coefficients (see Eqs. (29)-(31)). As a point of interest, we consider the lowest approximation
partition function having the form
\begin{equation}
 \label{int}
 Z_{V}^{0}=\frac{1}{2\pi i}\oint\frac{dz}{z}e^{|V|R(c(z))}.
\end{equation}
In the case $V\longrightarrow\infty$ integral (\ref{int}) could be
done using the steepest descent method
\begin{equation}
\label{partfun}
Z_{V}^{0}=\frac{1}{2\pi
i}\oint\frac{dz}{z}e^{|V|R(c(z))}\Longrightarrow\frac{1}{2\pi
i}\oint \frac{dc}{c}e^{|V|R(c_{m})}\Longrightarrow e^{|V|R(c_{m})}
\end{equation}
The point  of global maximum $c_{m}$ is defined by the equation
\begin{equation}
\label{paspoint}
\frac{d}{dc}R(c)=I_{1}(c)+I_{2}(c)+I_{3}(c)+N,
\end{equation}
The explicit form of functions $I_{1}(c),\ I_{2}(c),\ I_{3}(c),\
N$ reads
\begin{eqnarray}
\nonumber
&& I_{1}(c)=\int_{0}^{\infty}\frac{dk}{(2\pi)^{2}}k^{2}\left[
\frac{1}{\Delta(k)\tilde{V_{1}}(0)\left(
\frac{1}{\tilde{V_{1}}(k)}-\frac{1}{\tilde{V_{2}}(k)}\right)}
\right],\\
\nonumber
&& I_{2}(c)=\int_{0}^{\infty}\frac{dk}{(2\pi)^{2}}\frac{k^{2}}{\tilde{V_{1}}(0)\Delta^{2}(k)}\left(
\frac{1}{\tilde{V_{2}}(k)}-\frac{1}{\tilde{V_{1}}(k)}\right)\left[\frac{1}{\tilde{V_{1}}(k)\tilde{V_{2}}(k)}\left(2+\frac{c}{\tilde{V_{1}}(0)}\left(
\tilde{V_{1}}(k)-\tilde{V_{2}}(k)\right)\right)+1 \right],\\
\nonumber
&& I_{3}(c)=\int_{0}^{\infty}\frac{dk}{(2\pi)^{2}}\frac{\beta
k^{2}}{2}\frac{n}{\Delta^{2}(k)\tilde{V_{1}}(0)}\left(
\frac{1}{\tilde{V_{2}}(k)}-\frac{1}{\tilde{V_{1}}(k)}\right)^{2},\\
\nonumber
&& N(c)=\frac{1}{\beta
V_{1}(0)}-\frac{n}{c}-n\left(1-\frac{\tilde{V_{2}}(0)}{\tilde{V_{1}}(0)}\right)-\frac{c}{\tilde{V_{1}^{2}}(0)\beta}\left(
1-\frac{\tilde{V_{2}}(0)}{\tilde{V_{1}}(0)}\right).
\end{eqnarray}
The final expression for $R(c)$ is given by
\begin{equation}\label{state}
R(c)=M_{1}(c)+M_{2}(c)+P(c)
\end{equation}
where
\begin{eqnarray}
\nonumber
&& M_{1}(c)=\int_{0}^{\infty}\frac{dk}{(2\pi)^{2}}k^{2}n\beta\left(\tilde{V_{1}}(k)-\tilde{V_{2}}(k)
\right)\left[
\frac{1}{\Delta(k)\tilde{V_{1}}(k)\tilde{V_{2}}(k)}-1\right],\\
\nonumber
&& M_{2}(c)=\int_{0}^{\infty}\frac{dk}{(2\pi)^{2}}k^{2}\left[
\ln{\frac{1}{\Delta(k)\tilde{V_{1}}(k)\tilde{V_{2}}(k)}}+2-\frac{1}{\Delta(k)\tilde{V_{1}}(k)\tilde{V_{2}}(k)}\left[
2+\frac{c}{\tilde{V_{1}}(0)}\left(\tilde{V_{1}}(k)-\tilde{V_{2}}(k)
\right)\right]\right],\\
\nonumber
&& P(c)=n\left(\ln{n}-1-\ln{\frac{c}{\beta\tilde{V_{1}}(0)}}\right)+\left(1-\frac{\tilde{V_{2}}(0)}{\tilde{V_{1}}(0)}
\right)\left(\frac{c^{2}}{2\beta\tilde{V_{1}}(0)}-nc
\right)+\frac{c}{\beta\tilde{V_{1}}(0)}.
\end{eqnarray}
 All other thermodynamic functions may be found  from
$E(n,\beta)$ by the Maxwell relations in thermodynamics
\begin{equation}\label{davl}
P=- \left(\frac{\partial E}{\partial
V}\right)_{T}=\frac{1}{\beta}\left(\frac{\partial\ln{Z(n,\beta)}}{\partial
V}\right)_{T},\quad S=-\left( \frac{\partial E}{\partial
T}\right)_{V}.
\end{equation}
In particular, from above it follows
\begin{equation}
R(n,\beta)=P(n,\beta).
\end{equation}
 where $P$ is pressure of system.

\section{Summary}

 The developed procedure of computing the equation of state
 can be summarized as follows. First, we solve equation (\ref{paspoint}),
 and, second, the obtained roots are inserted in the equation  (\ref{state})
 which is the equation of state.
 Thus, the developed procedure permits one  to get the equation of
 state for  simple liquids, composed of particles  interacting via two-body
 potential with attractive and repulsive counterparts and having bound states.
 The application of this procedure to the simple liquid models with
 specific potentials is the subject of forthcoming article.
 
 We hope that the techniques presented in
this Letter can also be useful for other fields of computer
simulation, where the sign problem does occur. We
can contribute in this way to establish the auxiliary field
methodology as a standard tool for computation and this technique can also provide an interesting possibility to reduce the sign problem of other methods of computer simulation based on a functional integral approach.

 This work has been initiated by discussions with Prof. Garry V. Efimov to whom
 author is very indebted.

\section{ACKNOWLEDGMENTS}
We thank anonymous referees for valuable comments and suggestions.

\section{Appendix}

\subsection{Normal form of functional}
Let us introduce  normal form of functional with respect to Gauss
measure. Consider the equality:
\begin{equation}
\int\frac{D\phi}{C_{A}}e^{-\frac{1}{2}(\phi
D^{-1}\phi)+i(b\phi)}=e^{-\frac{1}{2}(bDb)}
\end{equation}
and identity:
\begin{equation}
\int\frac{D\phi}{C_{A}}e^{-\frac{1}{2}(\phi
D^{-1}\phi)}e^{i(b\phi)+\frac{1}{2}(bDb)}\equiv 1
\end{equation}
The normal form of functional with respect to Gauss measure with
Green function $D$ will understand multiplication
\begin{equation}\label{equality}
\vdots e^{i(b\phi)}\vdots_{D}\equiv e^{i(b\phi)+\frac{1}{2}(bDb)}.
\end{equation}
This definition is identity and valid with any b, therefore
 expanding  both hands of Eq.
~\ref{equality}  with respect to $b$ we obtain
\begin{displaymath}
\vdots \phi(x)\vdots_{D}=\phi(x),
\end{displaymath}
\begin{displaymath}
\vdots
\phi(x_{1})\phi(x_{2})\vdots_{D}=\phi(x_{1})\phi(x_{2})-D(x_{1},x_{2}),
\end{displaymath}
\begin{displaymath}
\vdots
\phi(x_{1})\phi(x_{2})\phi(x_{3})\vdots_{D}=\phi(x_{1})\phi(x_{2})\phi(x_{3})-D(x_{1},x_{2})\phi(x_{3})-D(x_{2},x_{3})\phi(x_{1})
-D(x_{3},x_{1})\phi(x_{2}),
\end{displaymath}
\begin{displaymath}
\ldots\ldots\ldots\ldots\ldots\ldots\ldots\ldots\ldots\ldots
\end{displaymath}
We can use the functional in normal form as:
\begin{equation}
\int d\sigma_{\phi,D}\vdots e^{i(b\phi)}\vdots_{D}\equiv 1
\end{equation}
\begin{displaymath}
\int d\sigma_{\phi,D}\vdots
\phi(x_{1}\ldots\ldots\phi(x_{n}))\vdots_{D}\equiv 0
\end{displaymath}
In particular:
\begin{displaymath}
\vdots e^{i(b\phi)}\vdots_{D}=e^{\frac{1}{2}(b[D-B]b)}\vdots
e^{i(b\phi)} \vdots_{B}
\end{displaymath}
 where
\begin{equation}\label{gauss}
I(g)=\int\frac{D\phi}{C_{A}}e^{-\frac{1}{2}(\phi
A^{-1}\phi)+gW[\phi]}=\int d\sigma_{\phi,A}e^{gW[\phi]}
\end{equation}
with normalization:
\begin{displaymath}
I(0)=\int\frac{D\phi}{C_{A}}e^{-\frac{1}{2}(\phi A^{-1}\phi)}=\int
d \sigma_{\phi,A}=1.
\end{displaymath}
Any  path integral over Gaussian measure for analytical $\phi$ can be written in normal
form with some Green function $D$:
\begin{equation}
W[\phi]=\int d\mu_{\eta}e^{i(b\phi)}=\int
d\mu_{\eta}e^{-\frac{1}{2}(\eta D\eta)}\vdots
e^{i(\eta\phi)}\vdots_{D}
\end{equation}
This path integral convenient to represent in the form:
\begin{displaymath}
W[\phi]=W_{0}+i(W_{1}\phi)-\frac{1}{2}\vdots \phi
W_{2}\phi\vdots_{D}+\vdots W_{I}[\phi]\vdots_{D}
\end{displaymath}
Where
\begin{displaymath}
W_{0}=\int d\sigma_{\eta}e^{-\frac{1}{2}(\eta D\eta)}
\end{displaymath}
\begin{displaymath}
i(W_{1}\phi)=\int d\sigma_{\eta}e^{-\frac{1}{2}(\eta
D\eta)}i(\eta\phi)
\end{displaymath}
\begin{displaymath}
\frac{1}{2}\vdots \phi W_{2}\phi\vdots_{D}=\frac{1}{2}\int
d\sigma_{\eta}e^{-\frac{1}{2}(\eta D\eta)}\vdots
(\eta\phi)(\eta\phi)\vdots
\end{displaymath}
\begin{displaymath}
\vdots W_{I}[\phi]\vdots_{D}=\int
d\sigma_{\eta}e^{-\frac{1}{2}(\eta D\eta)}\vdots
e_{2}^{i(\eta\phi)}\vdots_{D}=O(\vdots \phi^{3}\vdots_{D})
\end{displaymath}
\begin{displaymath}
e_{2}^{z}\equiv e^{z}-1-z-\frac{z^{2}}{2}.
\end{displaymath}

\subsection{Equations}

Let us consider path integral ~(\ref{gauss}) and make use of the following equivalent
 transformations. First, we shift the variable of integration
 $\phi(x)\longrightarrow\phi(x)+\xi(x)$. Second, 
  we write the functional of interactions
 in the normal form with respect to Gauss measure for new kernel $B^{-1}(x_{1},x_{2})$, we obtain
 \begin{eqnarray}
 \nonumber
I(g)&=&\int\frac{D\phi}{C_{A}}e^{-\frac{1}{2}(\phi A^{-1}\phi)-(\phi
A^{-1}\xi)-\frac{1}{2}(\xi A^{-1}\xi)+gW[\phi+\xi]}\\
 \nonumber
 &=&\frac{C_{B}}{C_{A}}\int\frac{D\phi}{C_{B}}e^{-\frac{1}{2}(\phi
B^{-1}\phi)}e^{-\frac{1}{2}(\phi[A^{-1}-B^{-1}]\phi)-(\phi
A^{-1}\xi)-\frac{1}{2}(\xi A^{-1}\xi)+g W[\phi+\xi]}\\\nonumber
&=&\frac{C_{B}}{C_{A}}\int d\sigma_{\phi,B}e^{-\frac{1}{2}\vdots
\phi[A^{-1}-B^{-1}]\phi\vdots_{B}-\frac{1}{2}([A^{-1}-B^{-1}]B)-(\phi
A^{-1}\xi)-\frac{1}{2}(\xi A^{-1}\xi)}\\
&=& e^{gW_{0}+ig(W_{I}\phi)-\frac{g}{2}\vdots \phi
W_{2}\phi\vdots_{B}+g\vdots W_{I}[\phi]\vdots_{B}}.
\end{eqnarray}
The major contribution to the functional integral gives Gauss
measure $d\sigma_{\phi,B}$, therefore the linear and quadratic
terms over the integration variable $\phi(x)$ should be absent.
Thus we obtain two equations
\begin{eqnarray}
 \mbox{\rm Equation I}: \ -(\phi A^{-1}\xi)+ig(W_{I\phi})=0,\\
 \mbox{\rm Equation II}: \ -\frac{1}{2}\vdots(\phi[A^{-1}-B^{-1}]\phi)
\vdots_{B}-\frac{g}{2}\vdots (\phi W_{2}\phi)\vdots_{B}=0.
\end{eqnarray}
For more details see \cite{Ef-91,Ef-95}.

\end{document}